\documentclass[twocolumn,secnumarabic,amssymb, nobibnotes, aps, prl, floatfix, superscriptaddress, nofootinbib,nobalancelastpage]{revtex4-2}

\usepackage{xcolor}
\usepackage{fancyhdr}
\usepackage{lastpage}
\usepackage{amsmath}
\usepackage{graphicx}
\usepackage[utf8]{inputenc}
\usepackage{bm}
\usepackage{epsfig}
\usepackage{hyperref}
\usepackage{orcidlink}

\usepackage{braket}
\usepackage{tensor}
\usepackage[version=4]{mhchem}
\usepackage{titlesec}
\usepackage{ifthen}
\usepackage{sidecap}
\usepackage{listings}
\usepackage[para,online,flushleft]{threeparttablex}
\usepackage{bm}

\usepackage[all]{hypcap}
\usepackage{float}
\usepackage[normalem]{ulem}

\usepackage{multibib}
\newcites{meth}{References}

\renewcommand{\figurename}{Figure}
\renewcommand{\tablename}{Table}

\setcitestyle{super}

\titleformat{\section}{\normalsize\raggedright\bfseries\fontsize{10}{12}}{\arabic{section}.}{1em}{}
\titleformat{\subsection}{\small\raggedright\bfseries}{\arabic{section}.}{1em}{}

\titlespacing\section{0pt}{12pt plus 4pt minus 2pt}{0pt plus 2pt minus 2pt}

\renewcommand{\vec}[1]{\bm{#1}}

\begin{document}

\begin{titlepage}
{\fontsize{26}{10} 
\textbf{\textcolor{black}{\flushleft  Matter radii  from interaction cross sections using  microscopic nuclear densities}}}\\

{
A. J. Smith$^{1,2,*}$~\orcidlink{0009-0009-7957-611X},
K. Godbey$^{1,2}$~\orcidlink{0000-0003-0622-3646},
C. Hebborn$^{3,1,2,\dagger}$~\orcidlink{0000-0002-0084-2561},
W. Nazarewicz$^{1,2}$~\orcidlink{0000-0002-8084-7425},
F. M. Nunes$^{1,2}$~\orcidlink{0000-0001-8765-3693} and
P.-G. Reinhard$^4$~\orcidlink{0000-0002-4505-1552}
}

{
\fontsize{6}{10}{
\selectfont

$^{1}$Facility for Rare Isotope Beams, Michigan State University, East Lansing, Michigan 48824, USA.

$^2$Department of Physics and Astronomy, Michigan State University, East Lansing, Michigan 48824, USA.

$^3$Université Paris-Saclay, CNRS/IN2P3, IJCLab, 91405 Orsay, France.

$^{4}$Institute for Theoretical Physics II, University of Erlangen, D-91058 Erlangen, Germany.

$^*$Corresponding author: smithan@frib.msu.edu

$^\dagger$Corresponding author: hebborn@ijclab.in2p3.fr
}
}

\end{titlepage}

{\bf 
Understanding how nuclear size evolves with the number of protons and neutrons tests our models of strongly interacting matter.  The nuclear charge (and proton) radii accessible through electromagnetic probes carry fundamental information on the saturation density and nuclear correlations. The radii of the neutron distribution are more difficult to measure, but they are important for our understanding of the isovector properties of nuclei that depend on the proton-to-neutron asymmetry, and on extended nucleonic matter in neutron stars.
Interaction cross sections offer one of the few direct experimental windows into the neutron radii of nuclei far from stability, but translating these measurements into reliable structural information requires an integrated theoretical framework that links structure and reactions with a rigorous treatment of uncertainty.
In this work, we compute interaction cross sections by using 
uncertainty-quantified proton and neutron distributions obtained in  
the self-consistent nuclear Density Functional Theory (DFT) with 
the Fayans energy density functional. The resulting densities are used in a modernized Glauber reaction framework, which features the refit of nucleon-nucleon profile functions.
Applying this pipeline to the existing data on the calcium isotopic chain, we find no evidence for the dramatic neutron swelling reported earlier. While focusing here on the Ca chain, the methodology proposed in this work is applicable to interaction cross section measurements across the nuclear chart and is well-suited for new experiments currently planned at leading rare isotope facilities.
}

Significant information exists on nuclear charge radii, primarily from 
studies using electromagnetic probes, including electron scattering, muonic-atom spectroscopy, and optical isotope-shift measurements \cite{Nortershauser2023}.
Charge radii data, especially along long isotopic chains, contain a wealth of information about bulk nuclear properties and local  nuclear shell structure, nucleonic superfluidity, and nuclear shapes, which can be traced back to properties of nuclear forces \cite{Ekstrom2015,Reinhard2016,Reinhard2017,Miller2019,Miyagi2025}.

 Compared to charge radii, matter and neutron radii deduced from experiments involving strongly-interacting probes \cite{Ozawa2023,Kanungo2023}  are  limited, plagued by model dependence, and less precise.
 Extracting 
 matter and neutron radii from measured interaction cross sections is then the most commonly used approach to get information on neutron radii of short-lived nuclei\cite{Ozawa2001,Tanaka2020,Tanihata1985,Suzuki1995,Ponnath2025} and, thus, on isovector properties of atomic nuclei, and on radii of neutron stars\cite{Horowitz2001,Fattoyev2018}.
The extraction of these radii from experimental interaction cross sections  is often done  using the modified optical limit approximation (MOL) within the 
Glauber model\cite{SuzukiBook,Glauber,Nagahisa2018}.
Typical  analyses relies on phase shifts  obtained by folding simple phenomenological nuclear  densities  with    parametrized profile functions for the nucleon-nucleon interactions. 
The extraction of nuclear radii is then performed by adjusting nuclear densities parameters to reproduce the measured interaction cross sections.
A similar procedure has recently been adopted in modeling ultra-relativistic heavy-ion collisions to extract nuclear densities~\cite{STAR2024}.
In both approaches, however, there are severe doubts as to the validity of the resulting inference on the nuclear structure properties if careful consideration is not given to the underlying statistical pipeline that connects structure inputs to reaction observables~\cite{Dobaczewski2025}.

This need for an integrated, end-to-end framework for the quantified analysis of reaction data has been further motivated in recent years by several studies that have observed discrepancies between theoretical predictions for nuclear radii and the corresponding radii extracted from reactions
\cite{Teixeira2022,Tanaka2020,Ponnath2025}.
As most theoretical uncertainties have not been quantified, it remains unclear whether these discrepancies arise from inaccuracies in the theoretical predictions of nuclear radii, in the reaction analysis used to extract radii from experimental data, or both.
Recently, efforts have been made~\cite{Horiuchi2025_short,Horiuchi2025_long} to assess the accuracy of the MOL by comparing it with calculations based on the full integration of the many-body wave function. Although applied only to light nuclei, this work shows that the MOL  is accurate for nucleus-nucleus reactions at energies above 100 MeV/nucleon. This suggests that the issue likely does not lie with the MOL itself, but with its inputs, the subsequent in-medium corrections\cite{Teixeira2022,SuzukiBook}, 
and/or an inconsistent treatment of the structure and reaction properties. 

In this work, we revisit the Glauber analysis by developing an integrated pipeline between the structure  and reaction calculations along isotopic chains. This is achieved  by directly integrating  densities predicted by a microscopic nuclear structure model, the Fayans energy density functional (EDF)~\cite{Reinhard2024}. To ensure consistency between the structure and reaction properties and to effectively account for in-medium effects, we calibrate the nucleon-nucleon profile functions using reaction measurements from a portion of the isotopic chain for an ensemble of predictions from the calibrated EDF.
We also carefully quantify the uncertainties associated with each parameter present in the reaction model and propagate them to interaction cross sections. This procedure allows us to  confront the cross section predictions with the experimental data directly.
To demonstrate  the performance of the improved Glauber methodology, we focus on the Ca chain. This isotopic chain is interesting because: (i) Excellent isotope-shift data exist and are accurately reproduced by the Fayans EDF\cite{Garcia_Ruiz2016,Miller2019}; (ii) The rise of matter radii for neutron-rich Ca isotopes, extracted from interaction cross sections \cite{Tanaka2020}, remains a puzzle for theory; (iii)
Interaction cross section data measured within the same campaign and setup exist\cite{Tanaka2020}; and (iv) Properties of calcium isotopes beyond $A=51$ are the subject of current interest at multiple rare-isotope facilities around the world.

We calculate
the interaction cross sections of  $^A$Ca projectiles on a  $^{12}$C target, using the MOL and neutron and proton densities calculated with Fayans EDF. Compared to previous works~\cite{Ozawa2001,Tanaka2020,Suzuki1995} that use a schematic two-parameter Fermi-type model of densities fitted to reproduce reaction data, our approach is designed to be more predictive and enforces consistency between the descriptions of the two reaction partners' structures. We use the recently developed Fayans EDF Fy(IVP3,0.9)\cite{Lalit2026} which employs isospin-variable pairing (IVP) terms.  We chose this specific model as it  has been carefully calibrated to many ground-state properties of nuclei, and its parametric uncertainties, associated with the calibration, have been quantified. Of particular interest for the current work, Fayans EDFs are capable of reproducing the complex pattern of charge radii along the Calcium chain \cite{Miller2019,Kortelainen2022}. In the following, we will not only use the optimal parametrization Fy(IVP3,0.9), but instead a representative ensemble  of parametrizations in its  1$\sigma$, 2$\sigma$, or 3$\sigma$ vicinity in order to assess the uncertainty of the cross section predictions  and perform model inference (see Methods).

The calculation of the interaction cross sections is done using the   MOL \cite{SuzukiBook,Glauber}, which has been shown to be more accurate than the standard optical limit approximation for analyzing interaction cross section~\cite{Abu-Ibrahim2000,Horiuchi2007,Nagahisa2018,Abu-Ibrahim2000_1,Takechi2009,Horiuchi2025_long}. Compared to other reaction theories, Glauber frameworks have the advantage that they rely on fewer  inputs and parameters, thanks to the fact that the radial dependence of the projectile-target interaction is obtained from by the  assumed nuclear densities. 
In contrast with previous works that use a profile function calibrated to nucleon-nucleon ($NN$) data\cite{Ray1979},  we adopt a different strategy: for each sample of projectile and target densities predicted by one parametrization in the vicinity of Fy(IVP3,0.9),  we recalibrate the parameters of the profile function to reproduce the interaction cross sections of the stable and long-lived  Calcium isotopes $^{42-48}$Ca (see Methods). This approach  allows us to effectively correct both for the  in-medium effects that are not treated in the MOL  as well as systematic inaccuracies associated with the  MOL approximation.  It also directly enforces consistency between the reaction treatment and the underlying EDF.
 Altogether, we deal with three sources of errors: from the uncertainty of the measured cross section, from the $NN$ profile function, and from the leeway of the EDF parametrization in their $\chi^2$ fits.

\begin{figure}
    \centering
    \includegraphics[width=\linewidth]{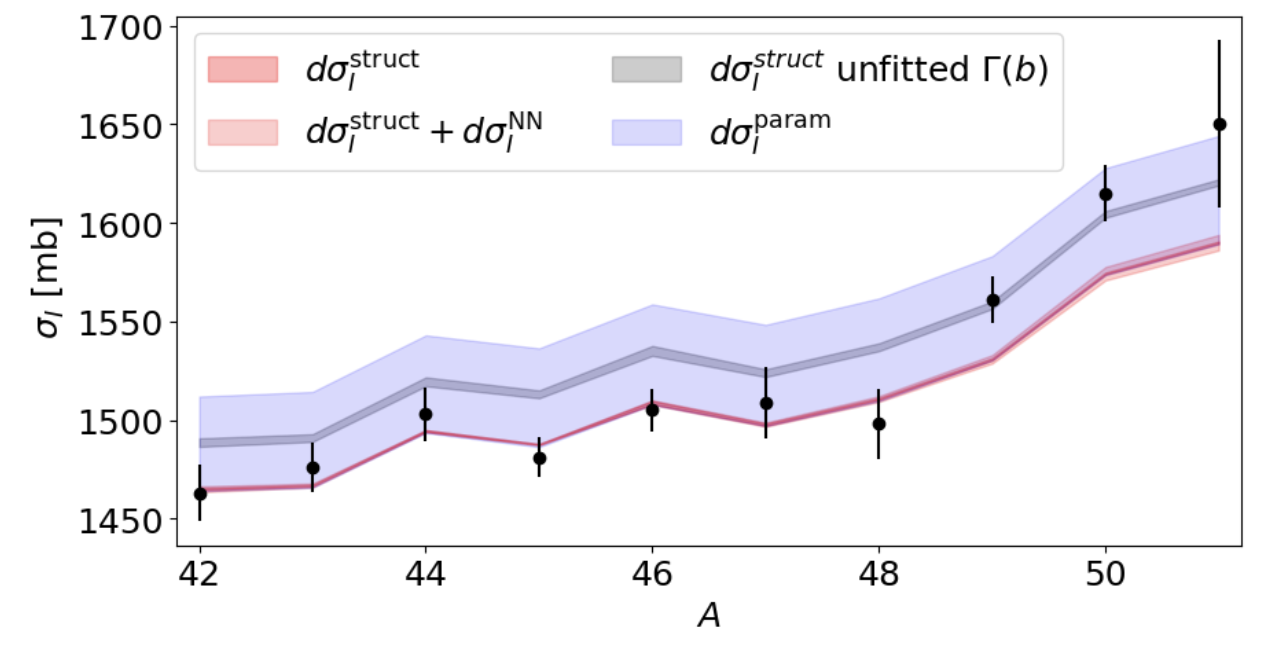}
    \caption{Comparison of experimental interaction cross sections\cite{Tanaka2020} (black
error bars, including both statistical and systematic errors) with theoretical predictions obtained using the MOL approach, densities predicted with the Fayans EDF, and profile function calibrated to $^{42-48}$Ca data. The dark-shaded red band corresponds to the  1$\sigma$ uncertainties $d\sigma^{struct}$ associated with the EDF calculations of nuclear densities. The gray band corresponds to the same calculations, except that we use a profile function from the literature\cite{Ray1979}. The light-shaded red band also includes the uncertainties associated with the $NN$ cross sections $d\sigma^{NN}$, that are used in the definition of the profile functions (see Methods). The uncertainties $d\sigma^{struct}$  and $d\sigma^{NN}$ are 
considered independent.  The blue bands correspond to parametric uncertainties related to the calibration of the profile function's free parameters.    }
    \label{fig:Sigma_r_MOL_Iso_a_not_0_param_uq}
\end{figure}

We first compare experimental interaction  cross sections $\sigma_I$ for $^{42-51}$Ca at 280~MeV/nucleon\cite{Tanaka2020} with our calculated values. Figure~\ref{fig:Sigma_r_MOL_Iso_a_not_0_param_uq} shows that the 1$\sigma$ uncertainties stemming only from the  calibration of the EDF parameters (red bands)  are negligible compared to the experimental errors. 
 The variations of the cross sections of $^{42-48}$Ca are  well reproduced, highlighting the accuracy of the Fayans EDF~\cite{Reinhard2024} to predict the odd-even staggering effect on nuclear densities\cite{deGroote2020}. For  neutron-rich Ca isotopes above $A=48$, our predictions reproduce a rise in the cross sections, but not as steep as observed experimentally.
Compared to calculations that use a profile function calibrated to $NN$ phase shift data of Ref.\cite{Ray1979}  (gray bands), our approach  reproduces the interaction cross section data more accurately. Moreover, the recalibration of the profile function for each Fayans EDF parametrization corrects for systematic deficiencies in the model. This is a way to account for the effects of model discrepancy.

We then quantify other sources of error in the reaction predictions, i.e., those associated with the evaluated $NN$ cross sections and with the calibration of the profile-function parameters.  Although the uncertainties on the neutron-neutron cross sections are large,  the uncertainties associated with $NN$ cross section (light-shaded red bands) are negligible for all Calcium isotopes. This is another benefit of the current approach: the uncertainties are reduced thanks to the recalibration of the profile function, which compensates for  inaccuracies of evaluated $NN$ cross sections. 
The uncertainties associated with the calibration of the profile function (blue bands in Fig.~\ref{fig:Sigma_r_MOL_Iso_a_not_0_param_uq}) are quantified using the covariance matrices evaluated at the best-fit parameter values and propagated to the cross sections. They are non-symmetric with respect to the best-fit parameters because of the strong nonlinearity of the MOL approach. These uncertainties are significantly larger than the experimental errors and the other theoretical uncertainties considered above. Nevertheless, they represent only 3\% of the interaction cross sections for all isotopes, which is considerably smaller than the parametric uncertainties associated with other reaction probes\cite{Lovell2018,King2018,Lovell2020,Catacora-Rios2023,Pruitt2024,Whitehead2022,Smith2024,Hebborn2023,Hebborn2024,Hebborn2023_1,Godbey2022,FrontiersHebbornNunes} that become larger for more isospin-asymmetric systems\cite{Smith2024}. The reduced uncertainties result from the direct incorporation of microscopic nuclear densities thereby fixing the radial dependence and consequently reducing the number of free parameters.

\begin{figure}
    \centering
    \includegraphics[width=1\linewidth]{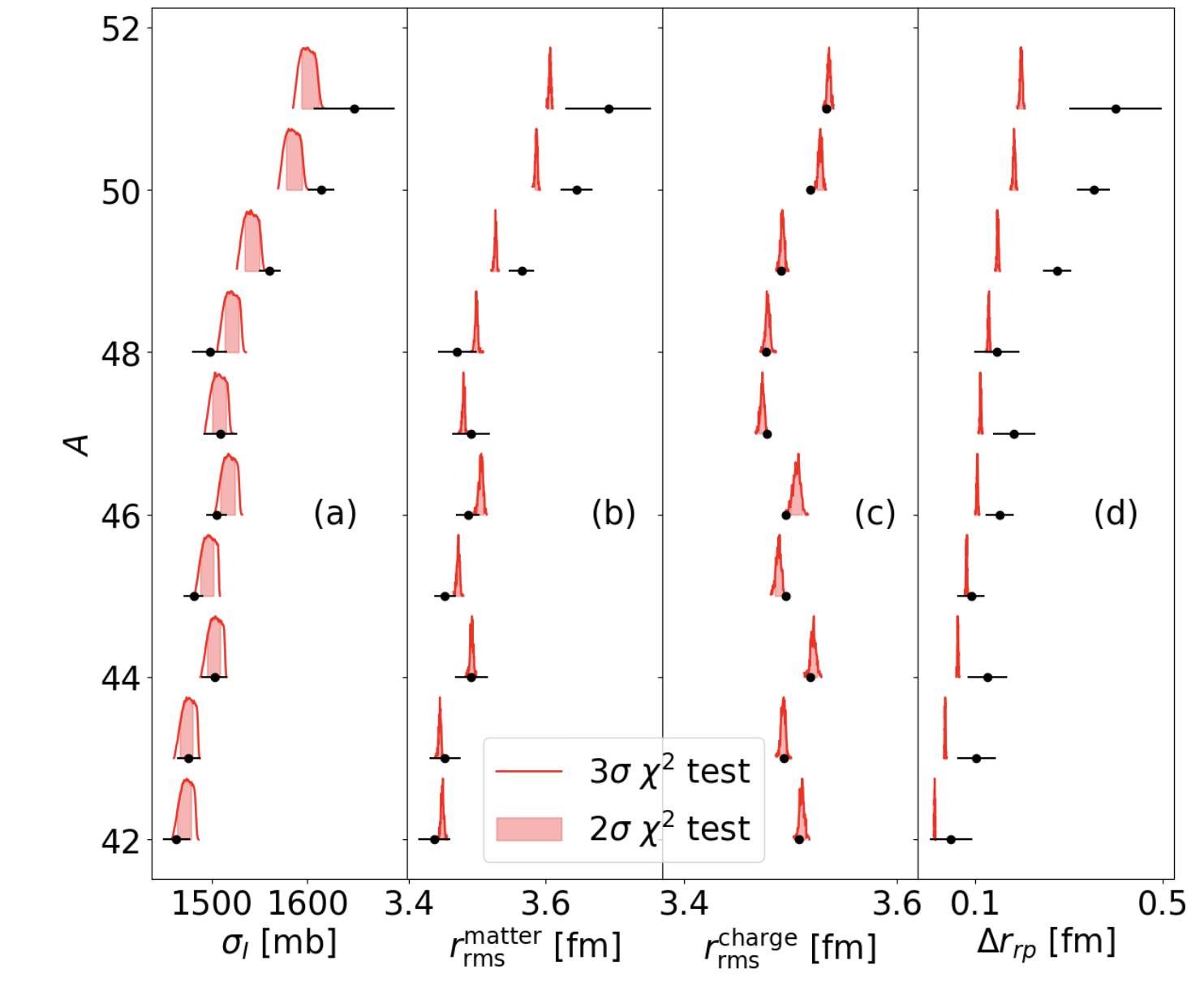}
    \caption{Theoretical predictions for (a) interaction cross sections, (b) matter radii, (c) charge   radii and (d) neutron skins (defined as differences between neutron and proton radii). The distribution delimited by the solid red line and the red-shaded distribution correspond respectively to the theoretical calculations that pass the $\chi^2$-test at the 3$\sigma$ and 2$\sigma$ confidence levels (no theoretical calculations pass the $\chi^2$-test at the 1$\sigma$ confidence level). This $\chi^2$-test is  performed on measurements of the interaction cross sections for the whole Calcium chain\cite{Tanaka2020} (black error bars in panel (a)). The error bars in panel (b) and (d) correspond to matter radii and neutron skins that were extracted in the initial analysis\cite{Tanaka2020} of these interactions cross sections data. The charge radii data are taken from\cite{Garcia_Ruiz2016}. }
    \label{fig:X_2_all_results}
\end{figure}

At first sight, after accounting for various sources of uncertainty, the model  predictions appear to be consistent with the experimental data, thereby removing the tension suggested in Ref.\cite{Tanaka2020}. Closer examination, however, indicates that the parametric uncertainties are strongly correlated among all nuclei: each set of profile-function parameters leads to cross sections that are shifted up and down in a similar way for all isotopes. 
 To account for these correlations, we perform a $\chi^2$-test comparing theoretical and experimental interaction cross sections across the entire Ca chain. The $\chi^2$-test is carried out for
 various samples of profile-function parameters, and the procedure is then repeated independently for all considered
 Fayans EDF parametrizations  from the given $n\sigma$ ensemble.
 This allows us to determine which Fayans EDF parametrizations are excluded at a given confidence level by the interaction cross-section data, while accounting for the parametric uncertainties associated with the profile functions. 
 Figure\,\ref{fig:X_2_all_results}(a) shows the  distributions that pass the $\chi^2$-test  at the $2\sigma$ and $3\sigma$ confidence levels for  the interaction cross sections, and Figures\,\ref{fig:X_2_all_results}(b-d) display    the corresponding  root-mean squared matter radii, charge radii, and neutron skins distributions. 

As expected,  the uncertainties due to the reaction process broaden the theoretical distributions of the interaction cross sections compared to those for the ground-state observables. The matter radii inferred from the reaction data~\cite{Tanaka2020}, shown in Figure\,\ref{fig:X_2_all_results}(b), do not include these errors, and thus have smaller uncertainties, leading to a more pronounced discrepancy between theory and experiments.
Remarkably, all DFT calculations pass the $\chi^2$-test on the interaction cross sections at the $2\sigma$ and $3\sigma$ confidence levels, demonstrating  the accuracy of the Fayans EDF to predict nuclear densities. The charge radii are also reproduced with high precision, although this is less surprising since some of these data were included in the calibration of the functional\cite{Reinhard2024,Lalit2026}.
The comparison with both charge radii and interaction cross section data gives confidence on the accuracy of the predicted neutron skin.   The discrepancy between our predicted neutron skin and those extracted in the original analysis of the interaction cross section measurements~\cite{Tanaka2020} further highlights the importance of an advanced treatment of reaction observables for achieving an accurate theory–experiment comparison.

Interestingly, none of our calculations pass the $\chi^2$-test for the interaction cross sections at the $1\sigma$ confidence level, due to the light tension between theory and experiment for the most neutron-rich Ca isotopes. 
More precise data on $^{51}$Ca  and new reaction data on  neutron-rich Ca isotopes with $A>51$ would help clarify this issue.

Finally, we have performed the same analysis but (i) considering the Fayans parametrization Fy($\Delta r$,HFB) that assumes isospin-independent pairing \cite{Reinhard2017}, (ii) changing the cross section data that were included in the calibration of the profile function parameters, i.e., we use  $^{42-46}$Ca and $^{42-51}$Ca cross section data,   (iii) considering the standard optical limit approximation instead of the MOL, and (iv) using matter densities instead of neutron and proton densities.  First, the analysis (i) showed that the discrepancy between theoretical and experimental cross sections is reduced when including the more realistic pairing term in Fy(IVP3,0.9). This confirms the need for isospin-dependent pairing to predict accurately the systematics of nuclear radii \cite{Garcia_Ruiz2016,Karthein2024,Reinhard2024}.
 The additional tests (ii-iv) confirm that our conclusions are robust with respect to the  reaction theory  employed and the dataset used in the calibration of the profile function.

This work establishes an integrated, uncertainty-quantified pipeline linking microscopic nuclear structure to interaction cross-section observables designed to meet the demands of current scientific programs at facilities such as FRIB and RIKEN, as well as next-generation rare-isotope experiments that will allow for the collection of many different data sets simultaneously.
By directly incorporating densities from the Fayans EDF and recalibrating the nucleon-nucleon profile functions consistently, the approach substantially reduces modeling uncertainties relative to conventional analyses, with parametric errors at the level of $3\%$.

Applied to the Ca chain, the framework finds no evidence for the dramatic neutron swelling reported in earlier analyses~\cite{Tanaka2020}, instead pointing to more modest neutron skin growth across the chain.
The light tension that remains for the most neutron-rich isotopes at the $2\sigma$ level motivates more precise measurements of $^{51}$Ca and measurements of heavier calcium isotopes, now becoming accessible at facilities such as FRIB and RIKEN.
Such data can be used to directly constrain EDF parametrizations, sharpen our understanding of isovector nuclear properties far from stability, and provide crucial information for understanding in-medium effects in intermediate energy reactions.

More broadly, the methodology presented here is immediately applicable to the analysis of interaction cross-section measurements across the nuclear chart. As high-statistics data arrive from upcoming campaigns at rare-isotope facilities, the pipeline's modular design enables rapid, statistically rigorous inference of matter and neutron radii from new measurements. This positions the framework as a practical tool for translating the coming generation of reaction experiments into reliable nuclear structure information.

\titleformat{\subsection}[runin]{\small\raggedright\bfseries}{\arabic{section}.}{1em}{}

\vfill\null
\section{\textcolor{blue}{METHODS}}

\setcounter{figure}{0}
\setcounter{table}{0}
\renewcommand{\figurename}{METHODS Figure}
\renewcommand{\tablename}{METHODS Table}
\renewcommand{\thefigure}{\arabic{figure}}
\renewcommand{\thetable}{\Roman{table}}
\renewcommand{\theHfigure}{METHODS Figure \thefigure}
\renewcommand{\theHtable}{METHODS Table \thetable}
 
\setcounter{equation}{0}

%\smallskip
\textbf{\textcolor{blue}{Nuclear models}} 
 \smallskip

The DFT  calculations were performed with the Fayans EDF. Density gradient terms in the pairing functional and surface energy allow a good description of the trend of nuclear charge radii, particularly in the Ca chain \cite{Garcia_Ruiz2016,Miller2019}. Here, we use the Fayans parametrization  with a finite-range pairing term Fy(IVP3,0.9) from Ref.\cite{Lalit2026}
that has been calibrated to a large dataset of ground-state properties in semi-magic nuclei \cite{Kluepfel_2009} complemented by  radius differences as summarized in Ref.\cite{Reinhard2024}.

The empirical calibration of the EDF implies some statistical uncertainties of the obtained parametrizations and of the corresponding predictions, following the rules of error propagation 
as discussed in Refs.~\cite{Kluepfel_2009,Dobaczewski2014}. The error propagation for the rather complex cross section calculations require an explicit sampling strategy.
Once the model has been calibrated, a set of 500 physically credible EDF samples (within $1\sigma$ or $2\sigma$ bands) 
are generated from the obtained covariance matrix assuming a multivariate gaussian distribution.
These samples are then used to obtain the ground state properties, including all relevant particle densities, of $^{12}$C and the Calcium isotopic chain which then feed into the reaction analysis pipeline. Averages, variances, and correlations can then be easily collected from the ensemble of results.

The interaction cross sections for the $^{42-51}$Ca projectiles on a $^{12}$C target at 280~MeV/nucleon  were calculated within the MOL of the Glauber framework\cite{SuzukiBook,Glauber}.
As done in previous works\cite{Abu-Ibrahim2000,Horiuchi2007,Nagahisa2018,Abu-Ibrahim2000_1,Takechi2009,Horiuchi2025_long,Tanaka2020,Ozawa2001,Suzuki1995}, we approximate interaction cross sections $\sigma_I$  by the reaction cross sections $\sigma_R$. The error associated with this approximation is estimated to be about 2\%,  based on data on $^{37}$Mg at 240~MeV/nucleon\cite{Takechi2014}. In the MOL formalism, the reaction cross sections   are obtained from the Glauber phase-shift function $\chi$ as
\begin{equation}
    \sigma_R =  \int   d^2\vec{b} \left( 1 - e^{ -2 \ \text{Im}  \chi(\vec {b})  } \right) 
\end{equation}
where $\vec b$ is the projectile's impact parameter vector. 

In this work, the phase-shift function are obtained from the  proton and neutron densities of the target  $T$ and projectile $P$ predicted by the DFT calculations (denoted $\rho^p_P$, $\rho^n_P$, $\rho^p_T$, and $\rho^n_T$) 
and profile functions for each $NN$ interaction (denoted $\Gamma_{pp}$, $\Gamma_{pn}=\Gamma_{np}$, and $\Gamma_{nn}$). The MOL phase-shift function reads
\begin{eqnarray}
&&\chi(\vec b) =  \sum_{ j,m =\{p,n\}}   \sum_{\substack{k,l = \{T,P\}\\ k\neq l } }\frac{{i}}{2}\nonumber \\
&&\times\int d^3 \vec r_k  \rho^m_k (\vec r_k) \left\{ 1 -  e^{ - \int  d^3 \vec r_l \rho^j_l(\vec r_l) \Gamma_{jm} (\vec b + \vec s_k -\vec s_l)} \right\},~~\strut
\end{eqnarray}
where $\vec r_P$ ($\vec r_T$) is the projectile (target) internal coordinate,
and $\vec s_P$ ($\vec s_T$) its transverse component.
We consider the parametrization of the profile functions proposed in Ref.\cite{Ray1979}
    \begin{equation}
        \Gamma_{jm}( b) = \frac{1 - i \alpha_{jm}}{ 4 \pi \beta_{jm}}  \sigma^{tot}_{jm}   e^{ - \frac{b^2}{ 2\beta_{jm}}},
        \label{profile_funct}
    \end{equation}
    where $\sigma^{tot}_{jm}$ are the $NN$ total cross sections evaluated at the beam energy,  $\alpha_{jm} $ and $ \beta_{jm}$ are parameters controlling the imaginary part of the scattering amplitude and the range  of the interaction. Since we consider a beam energy below the pion threshold, the unitarity of the $NN$ collision imposes\cite{Horiuchi2007} $\beta_{jm} = \frac{(1+ \alpha_{jm}^2)}{16 \pi} \sigma^{tot}_{jm}$.

    The gray band in Figure~\ref{fig:Sigma_r_MOL_Iso_a_not_0_param_uq} is obtained using the parameters of the profile function and the evaluated cross sections given in the original work\cite{Ray1979} where $\Gamma_{pp}(b) = \Gamma_{nn}(b)$.  Our theoretical calculations are obtained by recalibrating these profile function parameters and quantifying their uncertainties.\\

\textbf{\textcolor{blue}{Profile function calibration and uncertainty quantification}}

\textbf{Calibration of the profile function.}
Each $NN$ profile function~\eqref{profile_funct} depends on a total cross section $\sigma_{jm}^{tot}$ taken from Ref.\cite{Janout1972} and a single free parameter $\alpha_{jm}$.
For a given EDF parametrization, we calibrate $\alpha_{pp}$, $\alpha_{pn}$, and $\alpha_{nn}$ simultaneously by minimizing a $\chi^2$ function constructed from the measured interaction cross sections of $^{42-48}$Ca\cite{Tanaka2020}, using projectile and target densities predicted by that same EDF parametrization.
The minimization is performed via gradient descent.\\

\textbf{Propagating EDF parametric uncertainties.}
Uncertainties in the DFT calculations are propagated by drawing 500 physically credible samples of correlated $^{12}$C and $^{42-51}$Ca densities from the multivariate Gaussian defined by the EDF covariance matrix.
For each density sample, the profile function parameters $\alpha_{jm}$ are independently recalibrated, yielding 500 self-consistent (density, profile function) pairs. The resulting ensemble of 500 interaction cross-section calculations defines the $68\%$ confidence interval attributed to EDF parametric uncertainty.\\

\textbf{Propagating $NN$ cross-section uncertainties.}
The input cross sections $\sigma_{jm}^{tot}$ carry measurement or evaluation uncertainties\cite{Janout1972}, which we propagate via a finite-difference procedure. For each density sample, four profile-function fits are performed: fit (i) uses the central values $(\bar{\sigma}^{tot}_{nn}, \bar{\sigma}^{tot}_{pn}, \bar{\sigma}^{tot}_{pp})$, while fits (ii)--(iv) each shift one cross section by $+1\sigma$ in turn, leaving the other two at their central values.
The uncertainty contribution from each $NN$ channel is taken as the difference between the interaction cross sections obtained with fit (i) and the corresponding shifted fit; the three contributions are then added in quadrature to give a total $NN$ uncertainty for that density sample.
Averaging over all 500 density samples yields the $NN$ uncertainty band shown in Fig.~\ref{fig:Sigma_r_MOL_Iso_a_not_0_param_uq}.\\

\textbf{Propagating profile-function calibration uncertainties.}
For each of the 500 EDF density samples, we draw an additional 500 samples of $(\alpha_{pp}, \alpha_{pn}, \alpha_{nn})$ from a multivariate Gaussian centered on the best-fit values, with a covariance matrix evaluated at the minimum and rescaled by the best-fit reduced $\chi^2$, c.f. Ref.\cite{Dobaczewski2014}.
Combining EDF and profile-function samples yields 250,000 interaction cross-section calculations per isotope, from which the $68\%$ confidence intervals shown in Figure~\ref{fig:Sigma_r_MOL_Iso_a_not_0_param_uq} are derived.

\vspace{\baselineskip}

\textbf{\textcolor{blue}{Data availability}}

The data that support the plots within this paper and other findings of this study are available from the corresponding author upon reasonable request.\\

\textbf{\textcolor{blue}{Code availability}}

Our unpublished computer codes  used to generate results that are reported in the paper and related to its main claims will be made  available upon request to editors and reviewers.

\bibliographystyle{naturemag}
\bibliography{Biblio}

\vspace{\baselineskip}

\textbf{\textcolor{blue}{Acknowledgments}}

Useful discussions with Masaomi Tanaka are acknowledged. AJS thanks Pablo Giuliani, Claire Kopenhafer, and Grigor Sargsyan for interesting discussions.
This material is based upon research supported by the Chateaubriand Fellowship of the Office for Science \& Technology of the Embassy of France in the United States.
 It also received financial support from the CNRS through the AIQI-IN2P3 project.
 This material is based upon work supported by the U.S. Department of Energy under Award Numbers DE-SC0013365 and DE-SC0023688 (Office of Science, Office of Nuclear Physics), and DE-SC0023175 (Office of Science, NUCLEI SciDAC-5 collaboration). This work was also supported by the National Science Foundation CSSI program under award number OAC-2004601 (BAND Collaboration) and  the regional computing centre (RRZE) of the Friedrich-Alexander University Erlangen/N{\"u}rnberg.

\vspace{\baselineskip}
\textbf{\textcolor{blue}{Author contributions}}

AJS led the statistical analysis and performed the reaction calculations. PGR performed the EDF calibration. KG 
performed the DFT calculations. All authors discussed the results and contributed to the manuscript at different stages.

\vspace{\baselineskip}

\textbf{\textcolor{blue}{Competing interests}}

The authors declare that they have no competing financial interests.
\vspace{\baselineskip}

\textbf{\textcolor{blue}{Correspondence and requests for materials}}

Correspondence and requests for materials  should be addressed to A. J. Smith (email: smithan@frib.msu.edu) and C. Hebborn (email: hebborn@ijclab.in2p3.fr).

\end{document}